\documentclass{osa-article}

\journal{osajournal}

\articletype{Research Article}

\usepackage{microtype}
\usepackage[ruled]{algorithm2e}
\usepackage[colorinlistoftodos]{todonotes}
\usepackage{float}
\usepackage{amssymb}
\DeclareMathOperator*{\argmax}{argmax}
\usepackage{amsmath,amsfonts,bm}

\begin{document}


\title{Design space reparameterization enforces hard geometric constraints in inverse-designed nanophotonic devices}

\author{Mingkun Chen,\authormark{1} Jiaqi Jiang,\authormark{1} and Jonathan A. Fan\authormark{1,*}}

\address{\authormark{1}Department of Electrical Engineering, Stanford University, 348 Via Pueblo, Stanford, CA 94305, USA}

\email{\authormark{*}jonfan@stanford.edu} 



\begin{abstract}
Inverse design algorithms are the basis for realizing high-performance, freeform nanophotonic devices.  Current methods to enforce geometric constraints, such as practical fabrication constraints, are heuristic and not robust.  In this work, we show that hard geometric constraints can be imposed on inverse-designed devices by reparameterizing the design space itself.  Instead of evaluating and modifying devices in the physical device space, candidate device layouts are defined in a constraint-free latent space and mathematically transformed to the physical device space, which robustly imposes geometric constraints.  Modifications to the physical devices, specified by inverse design algorithms, are made to their latent space representations using backpropagation.  As a proof-of-concept demonstration, we apply reparameterization to enforce strict minimum feature size constraints in local and global topology optimizers for metagratings.  We anticipate that concepts in reparameterization will provide a general and meaningful platform to incorporate physics and physical constraints in any gradient-based optimizer, including machine learning-enabled global optimizers.
\end{abstract}






\section{Introduction}
Nanophotonic devices are capable of manipulating and guiding electromagnetic waves propagating in free space and on chip, and they have a broad range of applications in imaging, sensing, and optical communications.  Amongst the most effective methods to design these devices is gradient-based topology optimization \cite{lalau2013adjoint, piggott2015inverse, ReviewWerner}, which has been used to realize metagratings \cite{sell2017large}, metalenses \cite{phan2019high, bayati2020inverse, chung2020high}, and on-chip photonic devices \cite{molesky2018inverse} that utilize complex electromagnetic wave dynamics \cite{JianjiAnnalen} to achieve exceptional performance.  Topology optimization is performed by discretizing the device structure into small voxels, initializing each voxel with grayscale dielectric permittivity values $\varepsilon$, and then iteratively modifying these permittivity values in a manner that improves a Figure of Merit (FoM).  These modifications are based on gradient terms $\frac{\partial (\mbox{FoM})}{\partial \varepsilon}$ that can be computed for each pixel using the adjoint variables method \cite{hughes2018adjoint, fan2020freeform} or auto-differentiation \cite{minkov2020inverse, hughes2019forward}.  Topology optimization can be performed in the context of local optimization, in which local gradients are directly used to perform gradient descent on grayscale device structures \cite{fan2020freeform}, or global optimization with Global Topology Optimization Networks (GLOnets), in which local gradient calculations are combined with generative neural network training to perform global population-based optimization \cite{jiang2019simulator, jiang2019global}.

A critical issue concerning topology-optimized devices is the practical implementation of hard geometric constraints imposed by experimental considerations.  A particularly important geometric constraint is minimum feature size (MFS), which arises due to limitations in lithography patterning resolution \cite{yablonovitch1998optical, joy1983spatial,vieu2000electron, liebmann2001tcad} and etching aspect ratio \cite{wu2010high, marty2005advanced,yeom2005maximum, ng2013ultra}. The imposition of MFS constraints also enables proper base patterns to be defined and used in algorithms that enforce fabrication robustness, in which base patterns are co-designed with their geometrically eroded and dilated forms \cite{wang2019robust, wang2011robust, schevenels2011robust}.  Without these constraints, topology-optimized devices often possess complex geometric shapes with very small feature sizes, making them difficult if not impossible to experimentally fabricate in a reliable manner.

Current methods to impose MFS constraints fall in one of three classes. The first is to set the device voxel dimensions or spacing between features to match the desired MFS \cite{shen2015integrated,frei2007geometry, khoram2020controlling}.  While this method works, it adds significant granularity to the device design space, limiting final device performance.  The second is to use regularization terms that penalize the FoM when the MFS constraint is violated \cite{su2018fully, vercruysse2019analytical, huang2019implementation}. While this technique will generally push devices towards regions of the design space that satisfy the desired constraints, it does not guarantee their enforcement. A third method is to optimize the device in an unconstrained manner and then use threshold filters to incorporate constraints \cite{phan2019high, hughes2018adjoint, jiang2019simulator,jensen2005topology, zhou2015minimum}.  This method can be applied during the iterative optimization process or after unconstrained optimization is performed, and while it has the potential to work well, it only works when optimized devices in the unconstrained space locally map onto high performance optima in the constrained space.  

\begin{figure}[h]
  \centering
  \includegraphics[width=\linewidth]{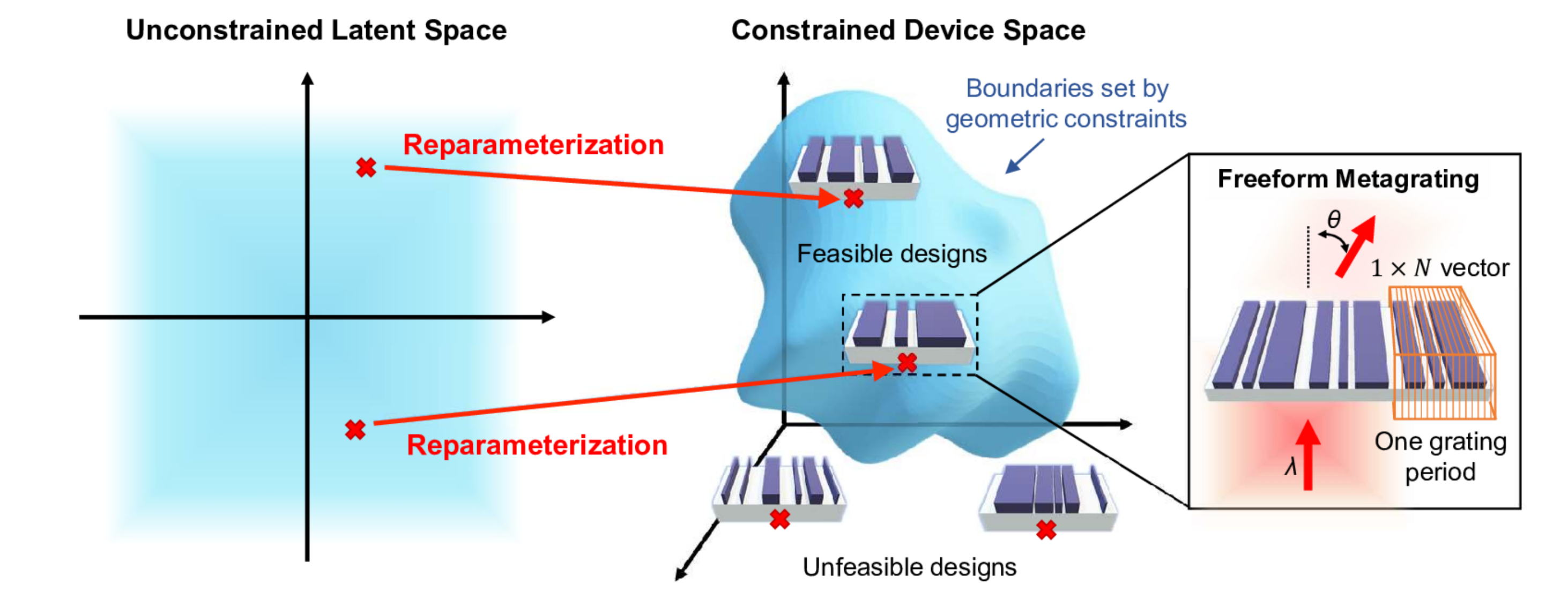}
  \caption{Overview of reparameterization for metagrating design.  The freeform metagratings comprise silicon nanoridges and are described by $1 \times N$ vectors.  To enforce hard physical constraints, mathematical transformations are used to map devices represented in an unconstrained latent space to a constrained device space. 
  }
  \label{fig:fig1}
\end{figure}

In this work, we propose to enforce strict geometric constraints within an inverse optimizer, while maintaining the fine pixel-level granularity of the physical design space, by reparameterizing the physical design space itself.  The idea, outlined in Fig. \ref{fig:fig1}, is to initially define devices in a latent space free of constraints and then transform these devices to a physical device space with hard constraints using analytic and differentiable mathematical transformations.  These devices are evaluated and optimized within this physical device space, and modifications to these devices are mapped back to their latent representations.  As a proof-of-concept, we demonstrate the implementation of reparameterization for the design of high efficiency metagratings using local optimization, based on the adjoint variables method, and GLOnets-based global optimization.  While we focus the application of reparameterization to the enforcement of feature size constraints, the concept provides a general formalism to incorporate physical information into black box inverse design tools.




\section{Problem setup}

Our model system is metagratings based on silicon nanoridges that deflect normally-incident light to the +1 diffraction order.  We have extensively studied metagratings in previous work using local \cite{sell2017large, sell2018ultra, Yang2017,YangOptLett} and global \cite{jiang2019simulator,jiang2019global} topology optimization, which allows us to thoroughly benchmark the results from this study with related state-of-the-art algorithms.  At all stages of optimization, an individual metagrating period is defined by a permittivity profile $\bm{\varepsilon} = \varepsilon(\emph{\textbf{x}})$, which denotes the material distribution at each voxel located at $\emph{\textbf{x}}$ within an individual grating period. $\varepsilon(\emph{\textbf{x}})$ is normalized so that 0 represents air and 1 represents silicon. The metagrating period has a width $L$ that is subdivided into $N = 256$ voxels, so that both $\emph{\textbf{x}}$ and $\bm{\varepsilon}$ are vectors with dimension $N$.  Our FoM is deflection efficiency, which is defined as the electromagnetic power deflected into the +1 diffraction order normalized to the total incident beam power.  We denote the deflection efficiency as $\mbox{Eff} = \mbox{Eff}(\bm{\varepsilon})$ and calculate it with the rigorous coupled-wave analysis (RCWA) electromagnetic solver Reticolo \cite{egorov2017genetically}.

The objective is to find a permittivity profile that maximizes the deflection efficiency, which can be cast as the following optimization:
\begin{align}
    & \mbox{maximize} & &\mbox{Eff}(\bm{\varepsilon}) \nonumber \\
    & \mbox{subject to} & & \bm{\varepsilon} \in \{0, 1\}^N \nonumber 
\end{align}
To define feature size constraints on the pixelized pattern $\varepsilon(\emph{\textbf{x}})$, we transform the pixelized pattern to a vector of discrete width values, $\emph{\textbf{w}}$, with which MFS constraints can be readily defined and incorporated.  For devices of fixed topology, each element in $\emph{\textbf{w}}$ represents ridge widths and air gap values.  We will see later in this study that this concept can generalize to devices of varying topology.  By reframing the pixelized patterns to a vector of geometric values, we reformulate the original optimization problem for devices of fixed topology to be:
\begin{align}
    &\mbox{maximize }  & &\mbox{Eff}(\emph{\textbf{w}}) \nonumber \\
    &\mbox{subject to} &&w_i \geq w_{min}, \qquad i = 1, 2, ..., M \label{eqw1} \\
    & & &\sum_{i=1}^{M} w_i = L \label{eqw2}
\end{align}
In Eq. (\ref{eqw1}), $w_{min}$ denotes the MFS, $w_i$ denotes the width of each structural feature, which can be a silicon ridge or air gap, and $M$ denotes the total number of silicon and air features.  Eq. (\ref{eqw2}) specifies that the total device width should be equal to the grating period $L$. 




\section{Reparameterization for local optimization}

\begin{figure}[h]
  \centering
  \includegraphics[width=\linewidth]{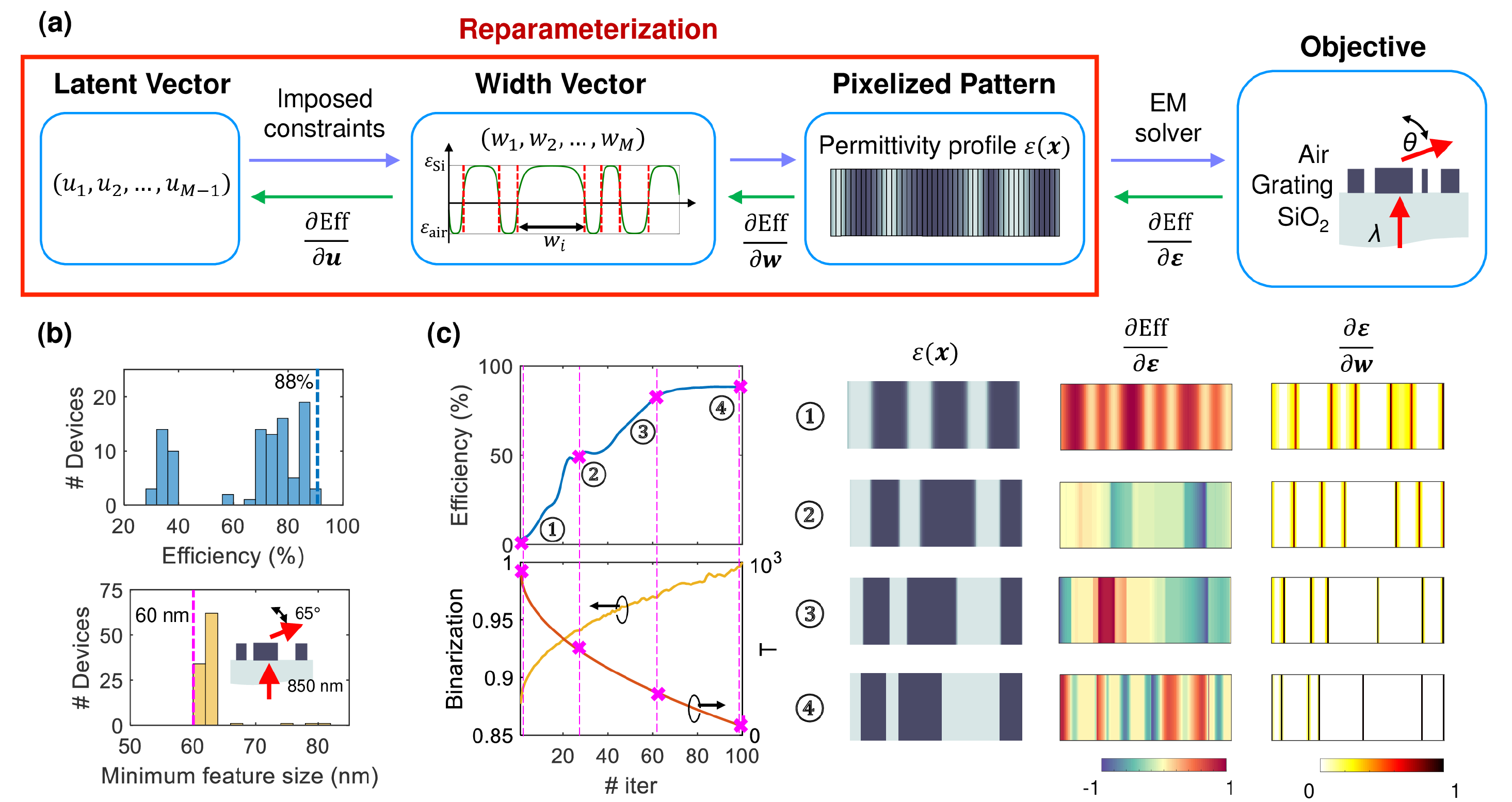}
  \caption{Reparameterized local optimization.\\
  (a) Flow chart of the reparameterized gradient-based local optimizer.  Latent vectors are transformed to physical devices, and gradients calculated for physical devices are backpropagated to update the latent vectors.  (b) Histograms of device efficiency and minimum feature size for 100 locally-optimized metagratings.  The inset shows the device operating parameters.  (c) Efficiency, binarization, and $T$ as a function of iteration number for a representative device optimization, together with device layouts and gradients for specific iterations.  Over the course of device optimization, the device efficiency increases, the pattern becomes more binary, and gradient contributions become more localized to the air-silicon boundaries.}
  \label{fig:fig2}
\end{figure}

To introduce the concept of reparameterization, we construct a reparameterized local optimizer for metagratings with fixed topology.  A flow chart of the design process is outlined in Fig. \ref{fig:fig2}(a) and the algorithm is summarized in Algorithm \ref{Algorithm1}. In the first step, a single device is randomly initialized with continuous latent vector values between -1 and 1.  This latent vector in the unconstrained latent device space is denoted by $\emph{\textbf{u}}$ and has elements $u_1,...u_{M-1}$ that can ultimately take any real value.  There are $M-1$ independent elements in this vector, as oppose to $M$ elements, due to the constraint set by Eq. (\ref{eqw2}).  Next, the latent vector is mapped to a width vector $\emph{\textbf{w}}$ through the following set of differentiable transformation functions:
\begin{align}
    &s_i=\mbox{sigmoid}(u_i) = \frac{e^{u_i}}{e^{u_i}+1}, \quad s_M=1, & &i = 1, ..., M-1 \label{equ2s} \\
    &t_1=s_1, \quad t_i=s_i \sqrt{1-\sum_{j=1}^{i-1} {t_j}^2 },  & &i = 2, ..., M\label{eqs2t} \\
    &w_i=\frac{t_i}{\sum_{j=1}^{M} t_j}(L-Mw_{min})+w_{min}, & & i = 1, ..., M\label{eqt2w}
\end{align}
We denote the transformation of $\emph{\textbf{u}}$ to $\emph{\textbf{w}}$ as $\emph{\textbf{w}} = f(\emph{\textbf{u}}, w_{min})$, and a more detailed derivation is provided in Supplement 1. In this manner, MFS constraints on the metagrating pattern, as defined in Eq. (\ref{eqw1}), have been imposed onto the width vector $\emph{\textbf{w}}$.

In the following step, $\emph{\textbf{w}}$ is mapped to a device in the pixelated device space, $\bm{\varepsilon}$, using a second set of analytic mathematical transformations.  Importantly, the resulting device is not a binary device consisting of silicon and air, but a grayscale device with a continuous range of dielectric permittivity values between silicon and air.  This relaxation of the device layout to a grayscale device ensures that local gradients can be properly calculated and used to improve device performance, which we will see later.  We define this mapping function $\bm{\varepsilon} = h(\emph{\textbf{x}}, \emph{\textbf{w}}, T)$ to be:
\begin{equation}\label{eqh}
    h(\emph{\textbf{x}}, \emph{\textbf{w}}, T) = \frac{1}{\exp(\pm\frac{(\tilde{\emph{\textbf{x}}}^{(i)})^2-(w_i/2)^2}{T}\cdot(\frac{L}{w_i M})^2)+1}, \qquad i=1,...,M
\end{equation}
The $\pm$ sign depends on the material of the $i^{th}$ structural feature and is $+$ for silicon and $-$ for air.  $T$ is a hyperparameter, analogous to a temperature term, which controls the degree of binarization of the pattern. $\tilde{\emph{\textbf{x}}}^{(i)}$ represents the position of pixels corresponding to the $i^{th}$ section, and they are defined such that the center of $w_i$ falls onto the position of $\tilde{x}^{(i)} = 0$.

With a physical device in hand, $\frac{\partial \mbox{Eff}}{\partial \bm{\varepsilon}}$ is evaluated using the adjoint variables method, in which forward and adjoint simulations are performed using the RCWA simulator.  This gradient term is then used to calculate $\frac{\partial \mbox{Eff}}{\partial \emph{\textbf{u}}}$ using backpropagation, which is based on the chain rule \cite{goodfellow_bengio_courville_undefined_2016}.  Backpropagation can be used in our algorithm because the entire reparameterization process can be described as a continuous and differentiable computation graph.  In practice, backpropagation is performed by programming our algorithm in PyTorch and using built-in auto-differentiation packages \cite{paszke2019pytorch}.  Finally, $\emph{\textbf{u}}$ is updated using the Adaptive Moment Estimation (Adam) algorithm, which is a variant of gradient descent \cite{kingma2014adam}, and the entire process is iteratively repeated.  

\begin{algorithm}[H]
\SetAlgoLined
\SetKwInOut{Parameter}{Parameters}
\Parameter{$w_{min}$, minimum feature size constraint. $M$, number of features. $\alpha$, step size. $\beta_1$ and $\beta_2$, momentum coefficients used in Adam.}
 initialization\;
 Sample $\emph{\textbf{u}} \sim \mathcal{U}^{M-1}(-1, 1)$\;
 \While{i < Total iterations}{
  $T=T_i$, set binarization level\;
  $\emph{\textbf{w}} = f(\emph{\textbf{u}}, w_{min})$, width vector in feature space\;
  $\bm{\varepsilon} = h(\emph{\textbf{x}}, \emph{\textbf{w}}^{(k)}, T)$, permittivity profile in device space\;
  $\mbox{Eff} = \mbox{Eff}(\bm{\varepsilon}), \quad \mathbf{g}_{\bm{\varepsilon}} = \frac{\partial \mbox{Eff}}{\partial \bm{\varepsilon}} \leftarrow$ forward and adjoint simulations\;
  $\mathbf{g}_{\emph{\textbf{u}}} = \frac{\partial \mbox{Eff}}{\partial \emph{\textbf{u}}} \leftarrow \mathbf{g}_{\bm{\varepsilon}}\cdot \frac{\partial \bm{\varepsilon}}{\partial \emph{\textbf{w}}} \cdot \frac{\partial \emph{\textbf{w}}}{\partial \emph{\textbf{u}}}$\;
  $\emph{\textbf{u}} \leftarrow \emph{\textbf{u}} + \alpha\cdot\mbox{Adam}(\emph{\textbf{u}}, \mathbf{g}_{\emph{\textbf{u}}})$\;
 }
 $\emph{\textbf{w}}^* = f(\emph{\textbf{u}}), \quad \bm{\varepsilon}^* = h(\emph{\textbf{x}}, \emph{\textbf{w}}^*, T), \quad \mbox{Eff}^* = \mbox{Eff}(\bm{\varepsilon}^*)$
 \caption{Reparameterization for local optimization}
 \label{Algorithm1}
\end{algorithm}

We use reparameterized local optimization to design metagratings that deflect normally-incident transverse-magnetic-polarized light at a wavelength of 850 nm to 65 degrees.  The MFS constraint is set to 60 nm, the topology is fixed to contain three silicon ridges $(M=6)$, and the device thickness is 325 nm. The refractive index of silicon is taken from Ref. \cite{green2008self} and only the real part of the index is used to simplify the design problem. A histogram of the deflection efficiencies of 100 different locally-optimized devices is presented in Fig. \ref{fig:fig2}(b) and shows a wide distribution of efficiency values, indicating that the design space is highly non-convex with many local optima.  The MFS distribution of these devices, also plotted in Fig. \ref{fig:fig2}(b), indicates that all devices have feature sizes larger than 60 nm, demonstrating that reparameterization enforces hard geometric constraints.

A more detailed analysis of a representative device from the histogram is shown in Fig. \ref{fig:fig2}(c).  The initial device pattern is random and set to have a large $T$.  In the high $T$ regime, the device possesses substantial regions of grayscale dielectric material, and the degree of binarization, defined as $\sum_{j=1}^{N}|2{\varepsilon}(x_j)-1|/N$, is low (as the device becomes more binary, this term gets closer to 1).  In addition, $\frac{\partial \bm{\varepsilon}}{\partial \emph{\textbf{w}}}$ spans many voxels (Fig. \ref{fig:fig2}(c)), which translate to large $\frac{\partial \mbox{Eff}}{\partial \emph{\textbf{w}}} = \frac{\partial \mbox{Eff}}{\partial \bm{\varepsilon}} \cdot \frac{\partial \bm{\varepsilon}}{\partial \emph{\textbf{w}}}$ upon backpropagation.  These gradients allow relatively large changes to the device layout to be made each iteration in the early stages of optimization, while the algorithm is broadly searching for a local optimum.  It is noted that the initial magnitude of $T$ cannot be too large and must be judiciously chosen because the grayscale and binary design spaces are different, and these differences become more substantial as $T$ increases.  Sufficient correlation between these two design spaces is required for gradients within the grayscale design space to reliably improve the device over the course of binarization. As the local optimizer evolves, $T$ is gradually reduced and the device becomes more binarized.  At this stage, $\frac{\partial \bm{\varepsilon}}{\partial \emph{\textbf{w}}}$ becomes more localized to the air-silicon interfaces  and the gradients provide fine-tuning of the device layout, in a manner akin to conventional boundary optimization.  Upon the completion of optimization, $T$ is decreased to zero and the device is fully binarized and contains only silicon and air material. 



\section{Reparameterized global optimization with fixed topology}





\begin{figure}[h]
  \centering
  \includegraphics[width=\linewidth]{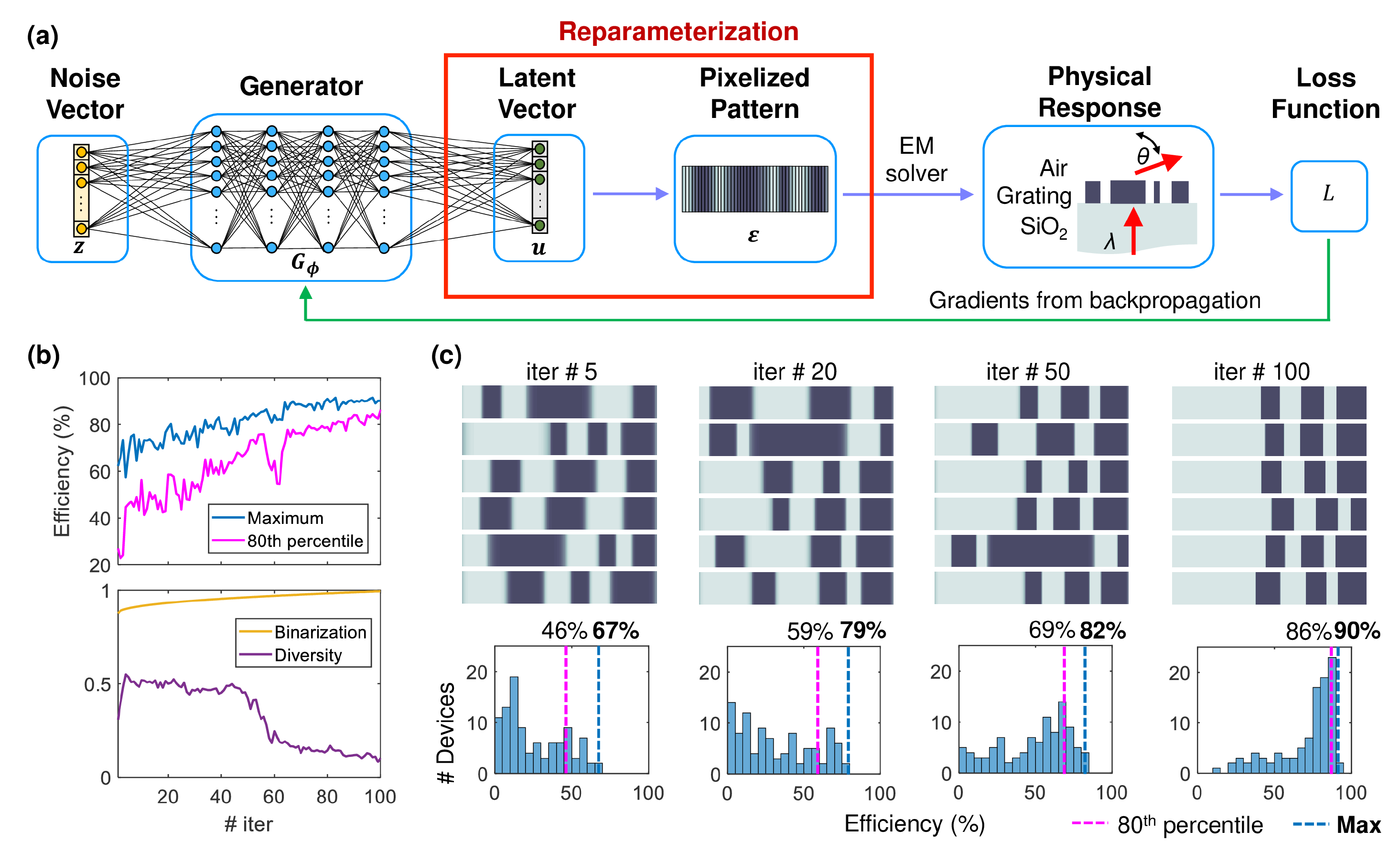}
  \caption{Reparameterized GLOnets for fixed topology devices.\\
  (a) Flow chart of reparameterized GLOnets.  A distribution of latent vectors are produced by a generative neural network and reparameterized to constrained physical devices.  Network weights are iteratively updated to minimize the loss function using backpropagation.  (b) Efficiency, diversity, and binarization as a function of iteration number for a representative GLOnets training run. (c) Histograms of generated device efficiencies and images of representative physical devices as a function of iteration number.  Over the course of optimization, the devices become more binary and the distribution of generated devices narrows and shifts to high efficiency values.}
  \label{fig:fig3}
\end{figure}

In this section, we show that reparameterization can be applied to GLOnets to enable population-based global optimization of fixed-topology devices with hard geometric constraints.  The flow chart and algorithm for reparameterized GLOnets are summarized in Fig. \ref{fig:fig3}(a) and Algorithm \ref{Algorithm2}, respectively.  Reparameterized GLOnets involve the iterative generation of devices from the network, evaluation of a loss function, and update of network weights that minimize the loss function.   Unlike the original GLOnets architecture, in which a generative neural network produces a distribution of physical devices \cite{jiang2019simulator}, the generator in the reparameterized GLOnets generates a distribution of latent vectors $\emph{\textbf{u}}$ from uniformly distributed noise vectors  $\emph{\textbf{z}} \in \mathcal{U}^{M-1}(-1, 1)$.  This nonlinear mapping from $\emph{\textbf{z}}$ to $\emph{\textbf{u}}$ is performed using a series of fully-connected layers: $\emph{\textbf{u}} = G_{\phi}(\emph{\textbf{z}})$, where $\phi$ are the network weights.   From a statistical perspective, the generator maps a uniform distribution of noise vectors to a probability distribution:
\begin{equation}\label{eqpp}
    G_\phi : \mathcal{U}^{M-1}(-1, 1) \mapsto P_{\phi}(\emph{\textbf{u}})
\end{equation}
where $P_{\phi}(\emph{\textbf{u}})$ denotes the probability of generating $\emph{\textbf{u}}$ in the latent device space.
The reparameterization process then follows, which maps $\emph{\textbf{u}}$ to the physical device pattern $\bm{\varepsilon}$ using the same formalism specified for the reparameterized local optimizer.  

The next step is the evaluation of the loss function using metrics calculated from the generated devices.  The loss function is engineered so that minimizing the loss function maximizes the probability that the neural network generates the optimal latent vector $\emph{\textbf{u}}^*$, which maps onto the globally optimized device $\bm{\varepsilon}^*$.  A formal derivation of the loss function is in Ref. \cite{jiang2019simulator}.  For reparameterized GLOnets, the loss function is defined to be:
\begin{equation}
    L(\bm{\varepsilon}, \mathbf{g}_{\bm{\varepsilon}}, \mbox{Eff}) = -\frac{1}{K}\sum_{k=1}^{K}  \frac{1}{\sigma} \exp{\left(\frac{\mbox{Eff}^{(k)}}{\sigma}\right)}\ \bm{\varepsilon}^{(k)}\cdot \mathbf{g}_{\bm{\varepsilon}}^{(k)}
    \label{Eqloss}
\end{equation}
The efficiencies, $\mbox{Eff}$, and efficiency gradients, $\mathbf{g}_{\bm{\varepsilon}} = \frac{\partial \mbox{Eff}}{\partial \bm{\varepsilon}}$, of the physical devices are calculated with forward and adjoint electromagnetic simulations.  $K$ is the batch size and $\sigma$ is a hyperparameter that biases network training towards devices possessing relatively high efficiencies and large gradients.  To minimize the loss function, backpropagation is performed to modify network weights $\phi$ in the generator.  As the mapping functions that link the noise vector, latent vector, and physical device pattern profile are continuous and differentiable, backpropagation is performed via the chain rule in a straightforward manner.  Upon the completion of network training, the network generates latent vectors that map to physical devices clustered around the global optimum. 

\begin{algorithm}[H]
\SetAlgoLined
\SetKwInOut{Parameter}{Parameters}
\Parameter{$w_{min}$, minimum feature size constraint. $M$, number of features. $K$, batch size. $\sigma$, loss function coefficient. $\alpha$, step size. $\beta_1$ and $\beta_2$, momentum coefficients used in Adam.}
 initialization\;
 \While{i < Total iterations}{
  $T:=T_i$, set binarization level\;
  Sample $\{\emph{\textbf{z}}^{(k)}\}_{k=1}^{K} \sim \mathcal{U}^{M-1}(-1, 1)$\;
  $\{ \emph{\textbf{u}}^{(k)} = G_{\phi}(\emph{\textbf{z}}^{(k)})\}_{k=1}^{K}$, latent vectors\;
  $\{ \emph{\textbf{w}}^{(k)} = f(\emph{\textbf{u}}^{(k)}, w_{min})\}_{k=1}^{K}$, width vectors\;
  $\{ \bm{\varepsilon}^{(k)} = h(\emph{\textbf{x}}, \emph{\textbf{w}}, T)\}_{k=1}^{K}$, permittivity profiles\;
  $\{\mbox{Eff}^{(k)}\}_{k=1}^{K}, \quad \{\mathbf{g}_{\bm{\varepsilon}}^{(k)}\}_{k=1}^{K} \leftarrow$ forward and adjoint simulations\;
  $g_\phi \leftarrow \nabla_{\phi}L( \{\bm{\varepsilon}^{(k)}\}_{k=1}^{K}, \{\mathbf{g}_{\bm{\varepsilon}}^{(k)}\}_{k=1}^{K}, \{\mbox{Eff}^{(k)}\}_{k=1}^{K}) $\;
  $\phi \leftarrow \phi - \alpha\cdot\mbox{Adam}(\phi, g_\phi)$\;
 }
 $\emph{\textbf{u}}^* \leftarrow \argmax_{ \emph{\textbf{u}} \in \{ \emph{\textbf{u}}^{(k)} | \emph{\textbf{u}}^{(k)} \sim P_{\phi^*} \}_{k=1}^{K}} \mbox{Eff(\emph{\textbf{u}})}$,\
 $\emph{\textbf{w}}^* = f(\emph{\textbf{u}}^*), \quad \bm{\varepsilon}^* = h(\emph{\textbf{x}}, \emph{\textbf{w}}^*, T), \quad \mbox{Eff}^* = \mbox{Eff}(\bm{\varepsilon}^*)$
 \caption{Fixed-topology reparameterized GLOnets}
 \label{Algorithm2}
\end{algorithm}



We first use reparameterized GLOnets to optimize metagratings with the same specifications as those designed by local optimization (Figs. \ref{fig:fig2}(b) and \ref{fig:fig2}(c)).  The results for a single network trained over the course of 100 iterations with a batch size of 100 devices is summarized in Figs. \ref{fig:fig3}(b) and \ref{fig:fig3}(c).  At the start of network training, the reparameterized generator has no knowledge of good metagrating designs and outputs a diversity of grayscale devices with modest efficiencies.
Device diversity for a given batch is computed as $\frac{1}{M}\sum_{i=1}^{M} \sqrt{\frac{\sum_{k=1}^{K}(w_i^{(k)}-\overline{w_i})^2}{K-1}}$. Over the course of network training, the best generated device and the overall device distribution shift towards higher efficiency regimes (Top panel, Fig. \ref{fig:fig3}(b)), and the devices become more binarized and less diverse (Bottom panel, Fig. \ref{fig:fig3}(b)). Upon the completion of network training, the best sampled device is simulated to have an efficiency of 90\%, which is higher than the best locally-optimized device in Fig. \ref{fig:fig2}(b).

The evolution of the device distribution over the course of network training is examined in more detail in Fig. \ref{fig:fig3}(c), which shows representative device layouts and performance histograms at different moments of optimization. In early iterations, the reparameterized generative network outputs a diverse distribution of grayscale devices, indicating that the network is broadly sampling the device design space in its search for good optima.  The performance histograms show mostly modest to low efficiency devices. At later iterations, the network has identified and converged to more narrow regions of the design space and outputs less diverse distributions of devices.  The final network produces binary devices that are clustered around the global optimum. 



\begin{figure}[h]
  \centering
  \includegraphics[width=\linewidth]{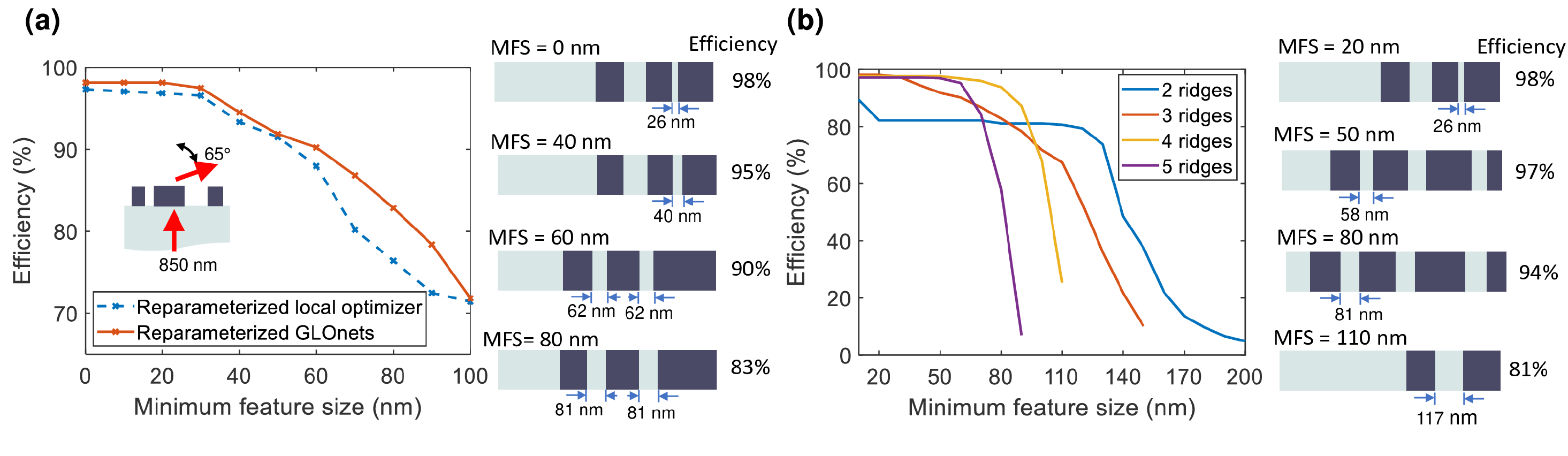}
  \caption{Performance of fixed-topology reparameterized GLOnets-designed devices for different minimum feature size (MFS) constraints.\\
  (a) Benchmark comparison of devices designed using reparameterized GLOnets and the reparameterized local optimizer, for devices comprising three silicon ridges and parameters matching those in Fig. \ref{fig:fig2}(b). (b) Plot of GLOnets-designed device efficiencies for differing device topologies.  The topology of the globally-optimized device strongly depends on the MFS constraint.}
  \label{fig:fig4}
\end{figure}

To further map out the metagrating design space, we optimize sets of devices with differing MFSs and topologies. We first present globally- and locally-optimized devices containing three silicon ridges, and the results are summarized in Fig. \ref{fig:fig4}(a).  Each data point in the reparameterized GLOnets curve represents the best device from a single trained network, while each data point in the reparameterized local optimizer curve represents the best of 100 individual optimizations.  All data points are calculated using the same number of RCWA simulations.  A comparison between these curves indicates that GLOnets consistently outperforms the local optimizer for nearly all MFSs, demonstrating its ability to effectively perform global optimization.  Both curves show a decrease in efficiency as the MFS constraint is increased, indicating that small features are advantageous in enhancing light diffraction efficiency.  The presence of small features in high performing devices is visualized in images of the device layouts, which show that the best globally-optimized devices possess at least one feature near or at the MFS length scale.  For MFS constraints smaller than 20 nm, the reparameterized GLOnets generate the same optimal device with a deflection efficiency of 98\%.  This plateauing follows because the unconstrained globally optimal device has a MFS of 26 nm.  

Globally-optimized efficiencies for devices with differing topologies are summarized in Fig. \ref{fig:fig4}(b), together with layouts of the best devices for a given MFS constraint. These data indicate that the topology and layout of the best device is a strong function of MFS.  For no constraints up to a 30 nm MFS, devices containing three silicon ridges produce the highest overall efficiencies.  Devices with four and five ridges also have efficiencies near the global optimum and have the benefit of larger feature sizes.  Above a MFS of 30 nm, the efficiencies of three ridge devices drop dramatically and the four and five ridge devices are optimal.  Above a MFS of 90 nm, two ridge devices have the highest overall efficiency.  

\section{Reparameterized global optimization with variable topology}

A proper global optimizer requires the ability to search for the proper device topology as well as the detailed layout for that topology.  Optimizers for fixed-topology devices can perform this task by parametrically sweeping across a wide range of topologies, as performed in Fig. \ref{fig:fig4}(b), but this route can be computationally intractable when scaling to complex systems.  In this section, we show that reparameterized GLOnets can combine concepts in boundary and topology optimization to globally optimize geometrically-constrained devices with variable topology.  

\begin{figure}[h]
  \centering
  \includegraphics[width=\linewidth]{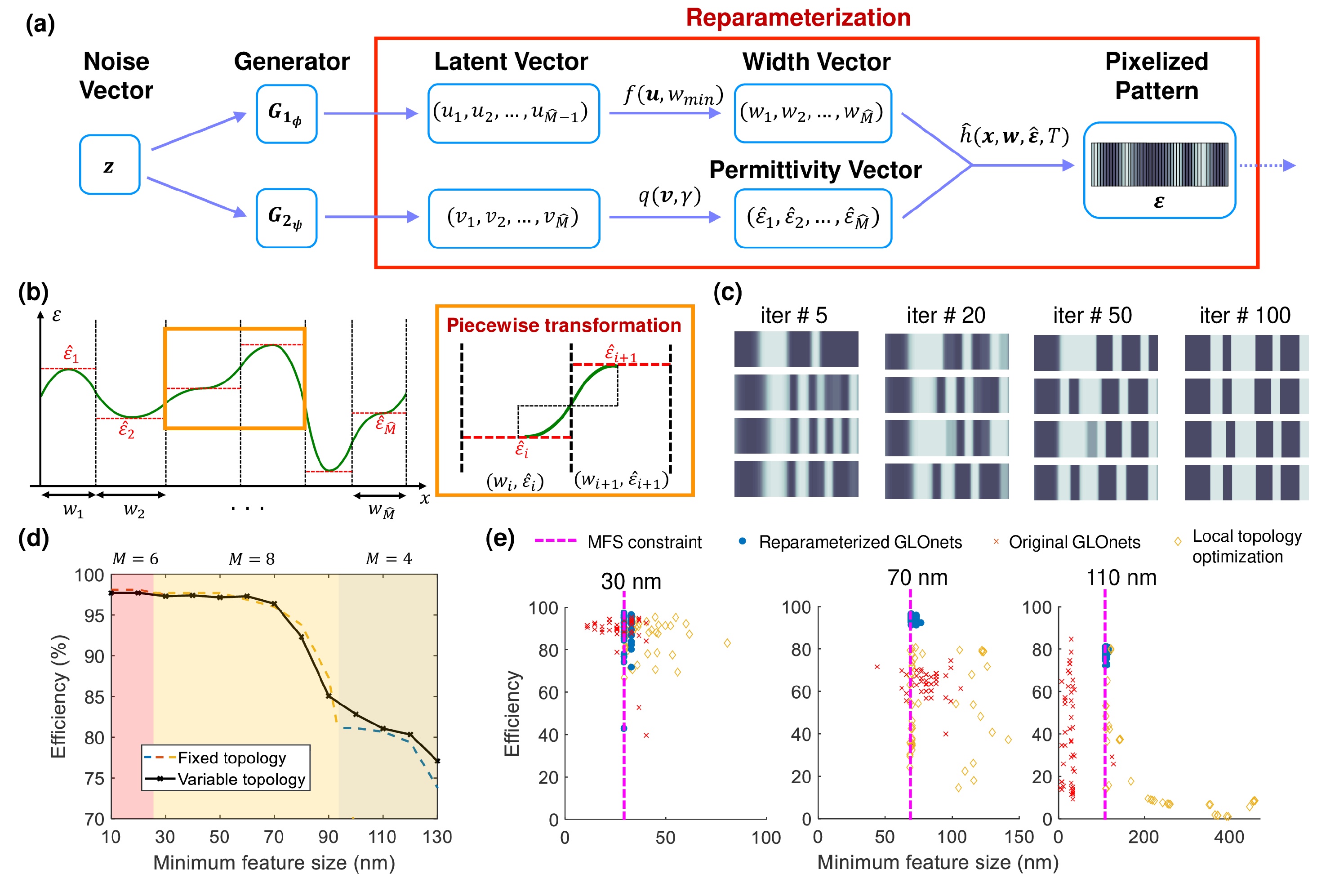}
  \caption{Reparameterized GLOnets for variable-topology metagratings.\\
  (a) Flow chart of the reparameterization process for variable-topology GLOnets.  Two separate generators are used to produce distributions of width and permittivity vectors, which are combined to produce constrained physical devices.  (b) Schematics of the piecewise transformation used for device reparameterization.  (c) Device layouts over the course of GLOnets training for a representative optimization run.  The devices are initially grayscale with varying topology and they converge to a narrow distribution of binary devices.  (d) Plot of device efficiencies as a function of the minimum feature size (MFS) constraint, designed using fixed-topology and variable-topology reparameterized GLOnets.  The parameters match those in Fig. \ref{fig:fig2}(b).  (e)  Scatter plots of device efficiency and MFS for devices designed using variable-topology reparameterized GLOnets, vanilla GLOnets, and adjoint-based local optimization, for three different MFS constraints.}
  \label{fig:fig5}
\end{figure}

The concept is outlined in Figs. \ref{fig:fig5}(a) and \ref{fig:fig5}(b).  As before with fixed-topology optimization, we subdivide the device permittivity profile into $\widehat{M}$ sections.  However, instead of assuming that a given section is either silicon or air, we define each section to have the parameters $(w_i,\widehat{\varepsilon}_i)$, where $w_i$ denotes the section width and $\widehat{\varepsilon}_i$ denotes the permittivity at the center of the section. A range of topologies can now arise because adjacent sections in final binarized devices can have either the same or different permittivity value.  The generalized permittivity profile of a device with our new scheme is calculated from the mapping function $\bm{\varepsilon} = \widehat{h}(\emph{\textbf{x}}, \emph{\textbf{w}}, \bm{\widehat{\varepsilon}}, T)$, where $\widehat{h}$ is a piece-wise transformation.  The permittivity profile of the entire structure is calculated by smoothly connecting the permittivity at the center of each section using a function similar to $h$ in Eq. (\ref{eqh})  (Figs. \ref{fig:fig5}(b) and S1 in Supplement 1). A detailed formalism of $\widehat{h}$ is discussed in Supplement 1.


In our newly configured GLOnets (Algorithm \ref{Algorithm3}), physical devices in this form are generated using two separate generative networks, one for section widths and the other for refractive indices now being optimized.  The widths $\emph{\textbf{w}}$ are produced the same way as with fixed-topology reparameterized GLOnets, where latent vectors $\emph{\textbf{u}}$ generated from $G_{1_\phi}(\emph{\textbf{z}})$ are reparameterized to $\emph{\textbf{w}}$ with MFS constraints. The permittivity vector $\bm{\widehat{\varepsilon}}  \in [0, 1]^{\widehat{M}}$, which determines the permittivity at the center of each section, results from latent vectors $\emph{\textbf{v}}  \in \mathbb{R}^{\widehat{M}}$ generated from $G_{2_\psi}(\emph{\textbf{z}})$, followed by the transformation $\bm{\widehat{\varepsilon}} = q(\emph{\textbf{v}}, \gamma)$:
\begin{equation}\label{eqq}
\widehat{\varepsilon}_i=\mbox{sigmoid}(\gamma v_i) = \frac{e^{\gamma v_i}}{e^{\gamma v_i}+1}, \quad i = 1, ..., \widehat{M}
\end{equation}
$\gamma$ is a tunable hyperparameter that controls the binarization of the permittivity vector $\bm{\widehat{\varepsilon}}$ and is analogous to the hyperparameter $T$ used to control the binarization of the device widths.  As with $T$, $\gamma$ is intially set to produce grayscale refractive index values and is manually increased in a gradual manner so that the final device possesses binary refractive index values of silicon or air.  $\widehat{M}$ specifies the total number of sections and sets the upper limit in the topological complexity of generated devices.  For example, if $\widehat{M}=10$, physical devices with up to five nanoridges can be generated.  


\begin{algorithm}[H]
\SetAlgoLined
\SetKwInOut{Parameter}{Parameters}
\Parameter{$w_{min}$, minimum feature size constraint. $\widehat{M}$, number of sections. $K$, batch size. $\sigma$, loss function coefficient. $\alpha$, step size. $\beta_1$ and $\beta_2$, momentum coefficients used in Adam.}
 initialization\;
 \While{i < Total iterations}{
  $T:=T_i, \gamma:=\gamma_i$, set binarization level\;
  Sample $\{\emph{\textbf{z}}^{(k)}\}_{k=1}^{K} \sim \mathcal{U}^{\widehat{M}-1}(-1, 1)$\;
  $\{ \emph{\textbf{v}}^{(k)} = G_{1_\phi}(\emph{\textbf{z}}^{(k)})\}_{k=1}^{K}, \{ \emph{\textbf{u}}^{(k)} = G_{2_\psi}(\emph{\textbf{z}}^{(k)})\}_{k=1}^{K}$, latent vectors\;
  $\{ \emph{\textbf{w}}^{(k)} = f(\emph{\textbf{u}}^{(k)}, w_{min})\}_{k=1}^{K}$, width vectors\;
  $\{ \bm{\widehat{\varepsilon}}^{(k)} = q(\emph{\textbf{v}}^{(k)},\gamma)\}_{k=1}^{K}$, permittivity vectors\;
  $\{ \bm{\varepsilon}^{(k)} = \widehat{h}(\emph{\textbf{x}}, \emph{\textbf{w}}^{(k)}, \bm{\widehat{\varepsilon}}^{(k)}, T)\}_{k=1}^{K}$, permittivity profiles\;
  $\{\mbox{Eff}^{(k)}\}_{k=1}^{K}, \quad \{\mathbf{g}_{\bm{\varepsilon}}^{(k)}\}_{k=1}^{K} \leftarrow$ forward and adjoint simulations\;
  $g_\phi \leftarrow \nabla_{\phi}L( \{\bm{\varepsilon}^{(k)}\}_{k=1}^{K}, \{\mathbf{g}_{\bm{\varepsilon}}^{(k)}\}_{k=1}^{K}, \{\mbox{Eff}^{(k)}\}_{k=1}^{K}) $\;
 $g_\psi \leftarrow \nabla_{\psi}L( \{\bm{\varepsilon}^{(k)}\}_{k=1}^{K}, \{\mathbf{g}_{\bm{\varepsilon}}^{(k)}\}_{k=1}^{K}, \{\mbox{Eff}^{(k)}\}_{k=1}^{K}) $\;
  $\phi \leftarrow \phi - \alpha\cdot\mbox{Adam}(\phi, g_\phi)$\;
  $\psi \leftarrow \psi - \alpha\cdot\mbox{Adam}(\psi, g_\psi)$\;
 }
 $(\emph{\textbf{u}}^*,\emph{\textbf{v}}^*) \leftarrow \argmax_{ (\emph{\textbf{u}},\emph{\textbf{v}}) \in \{ (\emph{\textbf{u}}^{(k)},\emph{\textbf{v}}^{(k)}) | \emph{\textbf{u}}^{(k)} \sim P_{\phi^*},\emph{\textbf{v}}^{(k)} \sim P_{\psi^*} \}_{k=1}^{K}} \mbox{Eff(\emph{\textbf{u}},\emph{\textbf{v}})}$,\
 
 $\emph{\textbf{w}}^* = f(\emph{\textbf{u}}^*), \quad \bm{\widehat{\varepsilon}}^* = f(\emph{\textbf{v}}^*,\gamma),\quad \bm{\varepsilon}^* = \widehat{h}(\emph{\textbf{x}}, \emph{\textbf{w}}^*, \bm{\widehat{\varepsilon}}^*,T), \quad \mbox{Eff}^* = \mbox{Eff}(\bm{\varepsilon}^*)$ 
 \caption{Variable-topology reparameterized GLOnets}
 \label{Algorithm3}
\end{algorithm}

The ability for variable-topology reparameterized GLOnets to search across different device topologies is first demonstrated for metagratings with the same operating parameters specified in Figs. \ref{fig:fig2}(b) and \ref{fig:fig3}(b).  
The network is trained using a total of 100 iterations with a batch size of 100, and device layouts at different stages of optimization are shown in Fig. \ref{fig:fig5}(c).  Near the beginning of the optimization at iteration \# 5, GLOnets generates a wide range of random devices with different topologies. As the training process evolves, the device distribution gradually converges to a narrow range of topologies, and by iteration \# 50, $G_{2_\psi}(\emph{\textbf{z}})$ has collapsed onto a single device topology.  GLOnets subsequently focuses only on boundary optimization to refine the four-ridge device layout.  At the end of training, the generated devices are clustered around the same global optimum as that achieved with Algorithm \ref{Algorithm2} for $M = 8$. 

To perform a more systematic benchmark analysis of our variable-topology reparameterized GLOnets, we design metagratings with parameters matching those in Fig. \ref{fig:fig4}(b) for a wide range of MFS constraints.  The results are plotted in Fig. \ref{fig:fig5}(d) together with the efficiencies of the best overall devices from Fig. \ref{fig:fig4}(b), and they show that variable-topology reparameterized GLOnets are able to search for the correct device topology for a given MFS constraint.  Furthermore, the best variable-topology devices (solid black line) largely follow the performance of the best fixed-topology devices (dashed color lines), indicating that this new variant of GLOnets can effectively perform reparameterized boundary optimization. 

We further benchmark variable-topology reparameterized GLOnets with two other methods.  The first is local adjoint-based optimization, where 200 unconstrained topology optimization iterations are first performed to produce binary devices from random grayscale devices, followed by 50 boundary optimization iterations to refine the binary device layouts and incorporate MFS constraints.  The second is the original GLOnets used in Ref. \cite{jiang2019simulator}, where a Gaussian filter is used at the network output in an attempt to impose a MFS constraint. The device designs are benchmarked by two criteria, deflection efficiency and the MFS in the physical device. 

Fig. \ref{fig:fig5}(e) shows scatter plots of the efficiencies and MFSs of devices designed using the three aforementioned techniques, for three MFS constraints.  Variable-topology reparameterized GLOnets generate devices that are clustered around high efficiency values and that strictly satisfy the MFS constraint. The original GLOnets are also able to find clusters of devices with high deflection efficiency.  However, these devices do not consistently satisfy the hard MFS constraints because the Gaussian filter is a soft constraint.  The combination of local topology and boundary optimization is able to produce devices that strictly satisfy the MFS constraint, but it does not guarantee high performance.  For small MFS constraints, the efficiencies of the best locally-optimized devices are comparable with those of the best reparameterized GLOnets devices, as the globally optimal unconstrained and constrained devices are in similar design space regions.  As such, enough instances of unconstrained local topology optimization will produce some devices near the globally optimal constrained device.  However, for large MFS constraints, the local optimization approach has difficulty in reliably finding the global optimum.  The reason is because the topology of devices with large MFS constraints can be very different from the unconstrained devices (Fig. \ref{fig:fig4}(b)), so that performing unconstrained topology optimization followed by local boundary optimization is not effective.  

\section{Robustness to fabrication imperfections}

\begin{figure}[h]
  \centering
  \includegraphics[width=\linewidth]{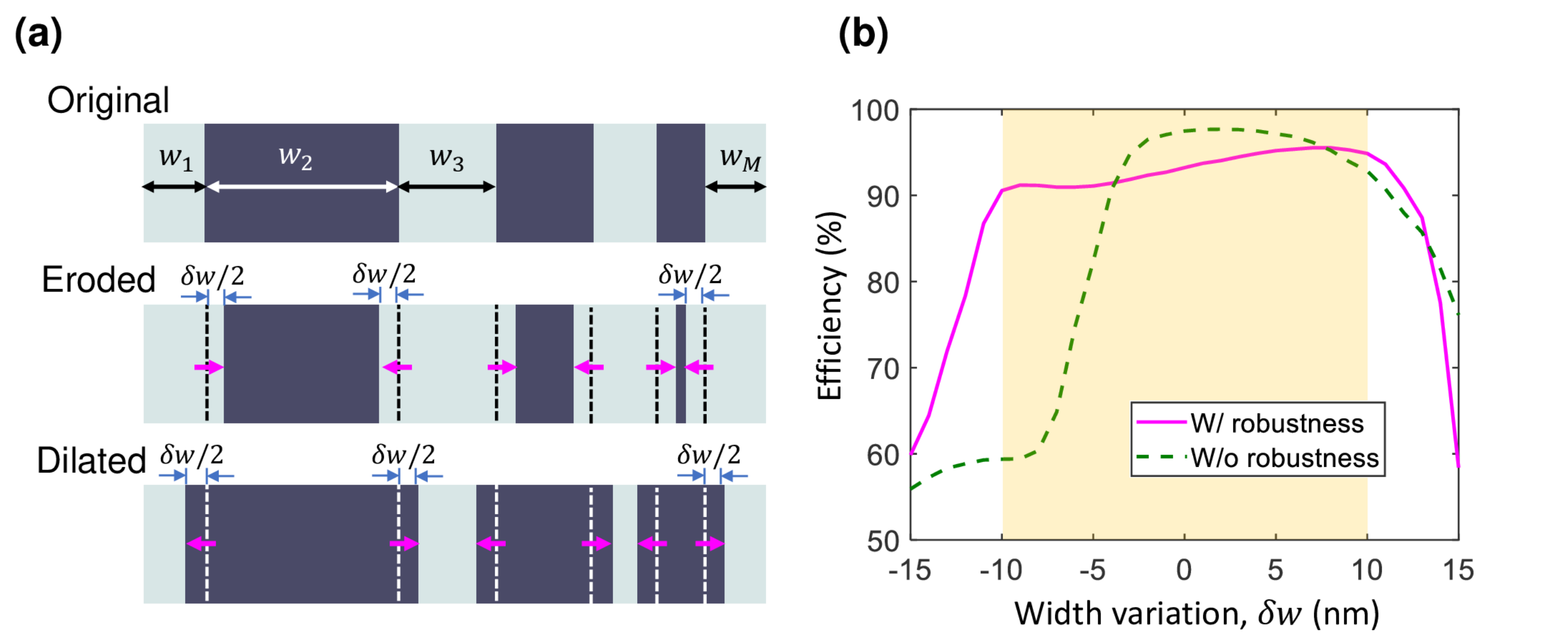}
  \caption{Robustness reparameterized GLOnets.\\
  (a) Fabrication imperfections are modeled by the erosion and dilation of device layouts, which is parameterized by $\delta w$. (b) Performance of devices globally optimized with and without robustness.  The device has a minimum feature size constraint of 30 nm, wavelength of 850 nm, and a deflection angle of 65 degrees, and the robust devices are designed to accommodate $\delta w = 10$ nm.}
  \label{fig:fig6}
\end{figure}

Devices that satisfy MFS constraints still do not guarantee good experimental performance as the device could be sensitive to other types of fabrication imperfections.
We show that robustness to fabrication imperfections can be readily incorporated into reparameterized GLOnets and focus this section on incorporating robustness in fixed-topology reparameterized GLOnets.  We leave the discussion of robust variable-topology reparameterized GLOnets to Supplement 1.  Robustness criteria are practically important due to geometric imperfections arising from all experimental fabrication processes.  A typical way to account for robustness, which we will implement here, is to consider the geometrically eroded and dilated versions of the devices in the FoM \cite{wang2019robust, wang2011robust, schevenels2011robust}. For our pixelized patterns $\varepsilon(\emph{\textbf{x}})$ generated from the width vector $\emph{\textbf{w}}$, eroded devices are realized by decreasing the silicon ridge widths by a width variation $\delta w$ and increasing the air gap widths by $\delta w$, and they are represented by width vector $\emph{\textbf{w}}_e$ (Fig. \ref{fig:fig6}(a)).  Dilated devices are realized by performing the opposite geometric transformations and are represented by width vector $\emph{\textbf{w}}_d$.  Our new FoM used in the GLOnets loss function is the weighted average deflection efficiency and is defined to be:
\begin{align}
    \mbox{Eff}_{\mbox{avg}} & = 0.5*\mbox{Eff}_{\mbox{original}} + 0.25*\mbox{Eff}_{\mbox{eroded}} + 0.25*\mbox{Eff}_{\mbox{dilated}} \label{eqre} \\
    & = 0.5*\mbox{Eff}(\emph{\textbf{w}}) + 0.25*\mbox{Eff}(\emph{\textbf{w}}_e) + 0.25*\mbox{Eff}(\emph{\textbf{w}}_d) \label{eqrw} 
    \end{align}

As a demonstration, we use fixed-topology reparameterized GLOnets with robustness to globally optimize a metagrating with a MFS constraint of 30 nm and $\delta w = 10$ nm.  The device operates with a wavelength of 850 nm and deflection angle of 65 degrees, and the topology is fixed to contain three silicon ridges.  The deflection efficiency of the globally-optimized device as a function of width variation is plotted in Fig. \ref{fig:fig6}(b) and shows that the device operates with a relatively high efficiency above 90\% for width variations within $\pm$\ 10 nm.  In comparison, the globally-optimized device without robustness has a higher overall peak efficiency when no width variation is present, but the efficiency rapidly decreases for negative width variations.  Variable-topology reparameterized GLOnets with robustness are as effective as fixed-topology reparameterized GLOnets with robustness at generating robust devices (Fig. S2 in Supplement 1).

\section{Conclusions and Future Directions}
We have shown that hard geometric constraints can be incorporated into gradient-based topology optimization frameworks using reparameterization.  This method involves mapping the constrained physical design space to an unconstrained latent device space in a manner that  naturally imposes constraints without impacting the granularity of the design space. We use these methods to incorporate MFS constraints into local and global optimizers for freeform metagratings, and benchmark calculations show that reparameterized GLOnets outperform gradient-based local optimizers and unconstrained GLOnets algorithms.


Future research directions include the extension of the reparameterization technique to the inverse design of more complex geometric structures, such as three-dimensional freeform metasurfaces.  The problem is challenging because these devices require constraints in both MFS and minimum radii of curvature.  One possible approach is to consider devices comprising ensembles of simple shapes with basic analytic descriptions, which provide fewer degrees of freedom compared to fully freeform layouts but which allow for clean calculations of feature size and curvature.  Another research direction is the use of reparameterization to account for more complex constraints, such as proximity error or side wall unevenness.  We anticipate that the use of continuous and differentiable functions to transform devices representations from latent to physical spaces will provide gradient- and neural network-based \cite{jiang2019free, wen2020robust, jiang2020deep} optimizers added functionality and facilitate the translation of high performance freefrom devices from theory to experiment.




\bibliography{refs}






\end{document}